\documentclass[fp,twocolumn]{jpsj3}

\usepackage[dvipdfmx]{xcolor}

\title{
Total-energy-assisted Tight-binding Method Based on Density Functional Theory
- Design Principles toward Transferability and Extrapolation }

\author{Takeo Fujiwara $^1$ \thanks{E-mail : u-tkfujiwr@g.ecc.u-tokyo.ac.jp},
        Yoshiro Nohara $^2$, 
and 
Susumu Yamamoto $^3$
}
\inst{$^1$ Department of Applied Physics, 
              The University of Tokyo, Bunkyo-ku, Tokyo 113-0033, Japan \\
$^2$ Exponential Design, Inc., Minato-ku, Tokyo 107-0062, Japan \\
$^3$ School of Computer Science, Tokyo University of Technology, 
                   Hachioji, Tokyo 192-0982, Japan \\
}		   
\date{Received: date / Accepted: date}

\abst{The previously proposed tight-binding method derived from the total energy has been extended 
within the local density approximation (TE-TB method; J. Phys. Soc. Jpn. 87, 064802 (2018)). 
Unlike the earlier two-parameter formulation, the present method introduces three parameters. 
These parameters explicitly capture local packing effects and improve transferability. 
Furthermore, based on physical considerations, several boundary conditions (inductive bias) are imposed on 
the functionals defining the tight-binding Hamiltonian and related energy functionals. 
The revised formalism is tested on crystalline silicon to assess the stability of the diamond structure, 
a monovacancy, and the (001) surface. 
These benchmark tests demonstrate the high transferability and reliability 
of the present three-parameter TE-TB design philosophy. }

\begin{document}
\maketitle
\section{Introduction} 
The tight-binding (TB) method is a versatile and widely used approach in quantum chemistry and condensed-matter physics. 
Owing to its computational efficiency, it is particularly well suited for large-scale molecular dynamics (MD) simulations. 
In addition, the TB method provides physical insight into structural, electronic, and chemical properties, 
and has been successfully applied to a wide range of materials and phenomena.
A notable extension is the combination of TB methods with the density functional theory (DFT) 
using molecular orbitals, known as the density functional tight-binding (DFTB) method.~\cite{Porezag_1995} 
The self-consistent charge (SCC) scheme was later introduced 
to account for charge-transfer effects among ions, leading to the SCC-DFTB.~\cite{M_Elstner_01,DFTB+}

In our previous paper~\cite{Fujiwara-Nishino-YamamotoECAL_2018}, 
we proposed a total-energy-assisted tight-binding method (TE-TB method) 
based on the total energy within the local density approximation (LDA), 
parametrized by the medium-range density and the effective contact radius. 
A major limitation of conventional TB and DFTB methods, including our previous work, 
is their limited transferability, which manifests itself as structure-dependent energy shifts 
and unstable extrapolation outside the fitted region. 
These difficulties originate from an implicit and insufficiently controlled description 
of local environments in the parameter space. 

In the present work, 
the TE-TB method is reformulated within a novel three-parameter framework to overcome these limitations. 
Section II briefly reviews the classification of the total energy, basis atomic orbitals, and the SCC method. 
Section III develops the three-parameter formalism. 
Section IV discusses the description of band structures and local energies within this formalism. 
Section V illustrates the design philosophy through applications to diamond-structured silicon, 
including its global stability, monovacancy, and (001) surface. 
Finally, Section VI presents the discussion and summary of the present work. 
In Appendix, the expressions of forces are summarized. 
The term `transferability' is used in the present paper in the sense of design control in parameter space, 
rather than extensive validation across multiple material systems.

\section{Short Review of TB-LDA}
\subsection{Total energy of LDA}\label{Total energy of LDA}

To formulate the tight-binding (TB) method based on the total energy expression 
within the local density approximation (LDA), 
the LDA total energy~\cite{Kohn-Sham} is taken as the starting point. 
Unless explicitly stated otherwise, all expressions are given in atomic units. 

The valence electron density $n({\mib r})$ is given as the sum of atomic electron densities 
$n({\mib r}) = \sum_a n_a({\mib r})$. 
Accordingly, the LDA total energy can be expressed as the sum of 
the band energy, the on-site energy, and the lattice energy:~\cite{Fujiwara-Nishino-YamamotoECAL_2018} 
\begin{subequations}
\begin{eqnarray}
 E_{\rm Total}^{\rm LDA} &=& E_{\rm band} + E_{\rm onsite} + E_{\rm lattice} ,      \label{Etotal-2}  \\
E_{\rm band}
 &=& \sum_i^{\rm occ} \varepsilon_i  , \label{Eband} \\
E_{\rm onsite}  
 &=&  E_{\rm xc}[n]-\int d{\mib r} n({\mib r})\mu_{\rm xc}({\mib r}) \nonumber\\ 
 && - \frac{1}{2} \sum_a \int d{\mib r}\int d{\mib r}^\prime  \frac{n_a({\mib r})n_a({\mib r}^\prime)}{|{\mib r}-{\mib r}^\prime|}
\nonumber \\
 &\equiv&  \sum_a E_{\rm onsite}^a ,   \label{onsite} \\
E_{\rm lattice} 
 &=& \sum_{\{a \ne b\}} \Big\{\frac{Z_aZ_b}{R_{ab}}
    -\int d{\mib r}\int d{\mib r}^\prime  \frac{n_a({\mib r})n_b({\mib r}^\prime)}{|{\mib r}-{\mib r}^\prime|} \Big\}  . \nonumber \\  
 & &  \label{Lattice-Coulomb}
\end{eqnarray}
\end{subequations}
Here, $\varepsilon_i$ is the eigenenergy of the $i$-th Kohn-Sham state. 
$E_{\rm xc}[n]$ and $\mu_{\rm xc}({\mib r})$ denote the exchange-correlation energy and potential, respectively. 
The last term represents the interaction between ions $a$ and $b$, separated by a distance $R_{ab}$. 
$Z_a$ is the effective atomic number of ion $a$, 
defined as the atomic number minus the number of core electrons. 
The summation $\sum_{\{a\ne b\}}$ runs over all distinct ion pairs. 

Once the atomic electron density $n_a({\mib r})$ is obtained within the present TB framework, 
the lattice (Coulomb) energy $E_{\rm lattice}$ can be evaluated. 
The lattice energy vanishes 
when the atomic charge densities do not overlap and each atom is electrically neutral.

\subsection{Atomic orbitals and TB Hamiltonian}
\subsubsection{Slater-type orbitals}
Slater-type orbitals (STOs)~\cite{J_Cerda_01} are adopted, 
since the overlap matrix elements can be evaluated analytically.~\cite{Mulliken1949,Zener1929}
An STO for an atom $a$, characterized by the principal, angular and magnetic quantum numbers 
($n,\ l,\ m$), is written using spherical harmonics $Y_l^m$ as 
\begin{eqnarray} 
  \phi_{a\mu}(\mathbf{r}) = R_{nl}^{(a)}(r)Y_l^m(\theta,\varphi) .
\end{eqnarray}
In the case of double-$\zeta$ representation,~\cite{J_Cerda_01} 
the radial function is given by 
\begin{eqnarray} 	 
  R_{nl}^{(a)}(r)  
     &=&  C_{(1)} \sqrt{\frac{(2\zeta_{l(1)}^{a})^{2n+1}}{(2n)!}}r^{n-1}\exp(-\zeta_{l(1)}^{a} r) \nonumber \\
     & &+ C_{(2)} \sqrt{\frac{(2\zeta_{l(2)}^{a})^{2n+1}}{(2n)!}}r^{n-1}\exp(-\zeta_{l(2)}^{a} r)   \ . \ \  
\end{eqnarray}
The ratio of $C_{(1)}/C_{(2)}$ of the coefficients is determined by the normalization condition.

Molecular orbitals (or band wavefunctions) are expressed as linear combinations of STOs 
\begin{eqnarray}
 \Psi_i= \sum_{a\mu}c_{a\mu}^{(i)}\phi_{a\mu}, \label{eq.MO}
\end{eqnarray}
where the coefficients $c_{a\mu}^{(i)}$ are obtained by solving the generalized eigenvalue equation 
\begin{equation}
   H\Psi_i = \varepsilon_i S \Psi_i ,
\end{equation}
with the overlap matrix element
\begin{equation}
  S_{a\mu,b\nu}=\langle \phi_{a\mu}|\phi_{b\nu}\rangle .
\end{equation}
The parameters $\zeta_{l(1)}$, $\zeta_{l(2)}$, $C_{(1)}$ and $C_{(2)}$ are determined 
so as to make the calculated band structure for a reference structure agree with that by the LDA calculation.

\subsubsection{Extended H\"uckel approximation}
The Hamiltonian form of the extended H\"uckel method~\cite{Hoffmann1,Wolfsberg-Helmholz} 
is adopted in the present formulation: 
\begin{subequations}
\begin{eqnarray}
 & & H_{\alpha, \alpha} = \varepsilon_\alpha 
= \varepsilon_{0\alpha}+\Delta_\alpha(\varrho,\eta), \label{eq.15b} \\
 & & H_{\alpha, \beta} = K\frac{\varepsilon_\alpha+\varepsilon_\beta}{2}S_{\alpha,\beta}, \ \   K=2.3 . \label{eq.15c}
\end{eqnarray}
\end{subequations}
Here $\alpha$ denotes an atomic orbital index. 
In this formulation, the local packing parameter $\eta$ is introduced 
in place of the constant energy shift employed in the previous TE-TB formulation.

\subsubsection{On-site energy}

The target on-site energy is defined as 
\begin{equation}
E_{\rm onsite}=E_{\rm total}^{\rm abinit}-E_{\rm band}-E_{\rm lattice},
\label{Def-trueEonsite}
\end{equation}
where $E_{\rm band}$ and $E_{\rm lattice}$ are obtained in the present TB method.
If the calculated on-site energy matches this target, 
the corresponding total energy coincides with the ABINIT results. 

\subsection{Self-consistent Charge (SCC) Scheme }
\subsubsection{Second-order perturbation theory}
The SCC approach proposed by Elstner et al.~\cite{M_Elstner_01} accounts for charge-transfer effects 
within second-order perturbation theory. 

In the SCC method, charge-transfer is described using the Mulliken charge $q_{a\mu}$ 
on orbital $\mu$ of atom $a$:
\begin{eqnarray}
 q_{a\mu} = 
    \frac{1}{2}\sum_i^{\rm occ} \sum_{b \mu \nu}
    \Big({c_{a\mu}^{(i)}}^* c_{b\nu}^{(i)} S_{a\mu,b\nu} 
        +{c_{b\nu}^{(i)}}^* c_{a\mu}^{(i)} S_{b\nu,a\mu} \Big). 
\end{eqnarray}

The total energy is then expressed as 
\begin{subequations}
\begin{eqnarray}
 && E^{\rm LDA}[n_0+\delta n] = E_0^{\rm TB}[n_0] +E_{\rm 2nd} , \label{eq.Hamiltonian1}\\
 && E_0^{\rm TB}[n_0] = \sum_i^{\rm occ}\langle \Psi_i|H_0|\Psi_i\rangle +E_{\rm onsite} +E_{\rm lattice} ,\ \ \ \ \label{eq.Hamiltonian2} \\
 && E_{\rm 2nd}= \frac{1}{2}\sum_{a\mu b\nu} \Delta q_{a\mu} \gamma_{a\mu, b\nu} \Delta q_{b\nu} . \label{eq.Hamiltonian3}
\end{eqnarray}
\end{subequations}
Here, $\delta n$ denotes the charge fluctuation,  
$H_0$ is the Kohn-Sham Hamiltonian 
and $\Psi_i$'s are the Kohn-Sham states corresponding to $E^{\rm LDA}[n_0+\delta n]$. 
$\gamma_{a\mu, b\nu}$ is the interaction between $a\mu$ and $b\nu$ orbitals, discussed in the next subsection. 

Depending on the local environment during an iteration step, 
the charge density may differ from its initial value. 
The deviation of the Mulliken charge is defined as~\cite{M_Elstner_01}
\begin{eqnarray}
\Delta q_{a\mu} = q_{a\mu}-q_{a\mu}^{(0)} ,
\end{eqnarray}
where $q_{a\mu}^{(0)}$ is the Mulliken charge on the $\mu$ orbital of atom $a$ of the electron density $n_0$.

\subsubsection{Variational formalism}
The variational formulation corresponding to Eqs.~(\ref{eq.Hamiltonian1})$\sim$(\ref{eq.Hamiltonian3})
is given by 
\begin{eqnarray}
& & \sum_\beta  \Big[\langle \phi_\alpha|H_0|\phi_\beta\rangle  \nonumber \\
& & \ \ \   + \frac{1}{2} S_{\alpha,\beta}  \sum_{\delta} (\gamma_{\alpha,\delta}+\gamma_{\beta,\delta})\Delta q_{\delta}
            - \varepsilon_i S_{\alpha,\beta} \Big] c_{\beta}^{(i)} =0,  \ \ \ \ \  \label{eq.Secular}
\end{eqnarray}
where $\alpha=(a, \mu)$, $\beta=(b, \nu)$, and $\delta=(c, \kappa)$ are used for notational simplicity.
The coefficients $c_{a\mu}^{(i)}$ define the Kohn-Sham states for $n=n_0+\delta n$. 
The {\it effective} Hamiltonian matrix elements are defined as 
\begin{subequations}
\begin{eqnarray}
H_{a\mu,b\nu} &=& H_{a\mu,b\nu}^{\rm 0-TB} +H_{a\mu,b\nu}^{\rm SCC},  \label{effective_H} \\
H_{a\mu,b\nu}^{\rm 0-TB}
&=& \langle \phi_{a\mu}|H_0|\phi_{b\nu}\rangle   ,                                            \label{Hamiltonian:TB-0} \\
H_{a\mu,b\nu}^{\rm SCC} 
 &=& \frac{1}{2}S_{a\mu,b\nu}\sum_{c\kappa} (\gamma_{a\mu, c\kappa}
                                            +\gamma_{b\nu, c\kappa})\Delta q_{c\kappa}  . \ \ \ \ \label{Hamiltonian:SCC}  
\end{eqnarray}
The first term (\ref{Hamiltonian:TB-0}) represents the usual TB expression for the LDA Hamiltonian $H _0$, 
while the second term (\ref{Hamiltonian:SCC}) accounts for the SCC contribution.

The diagonal term $\gamma_{a\mu,a\mu}$ can be estimated from the ionization and electron-affinity energies. 
Assuming that the interaction parameter $\gamma_{a\mu,b\nu}$ is independent of the orbital 
($\gamma_{a\mu,b\nu}=\gamma_{a,b}$), 
the diagonal term $\gamma_a \equiv \gamma_{a,a}$ and off-diagonal term $\gamma_{a,b}$ \ ($a \ne b$) 
can be expressed approximately as 
\begin{eqnarray}
&& \gamma_a=U_a, \ \ \ \\
&& \gamma_{a,b}=\frac{1}{R_{ab}}{\rm erf}\Big(\frac{\sqrt{\pi}}{2}\sqrt{\frac{2U_a^2U_b^2}{U_a^2+U_b^2} } R_{ab}\Big),  \label{eq.13e}
\end{eqnarray}
which satisfies (if the atoms $a$ and $b$ are the same element)
\[
 \lim_{R_{ab}\rightarrow 0} \gamma_{a,b}=\gamma_a .
\]
The expression (\ref{eq.13e}) is based on an approximation 
for a charge density around each atom as a spherically symmetric Gaussian function.

When the off-site contribution $\gamma_{a,b}$ of an atom pair $(R_{ab}\ne 0)$ can be neglected, 
the SCC contribution reduces to 
\begin{eqnarray} 
H_{a\mu,b \nu}^{\rm SCC} 
   =\frac{1}{2} S_{a\mu,b\nu}(\gamma_{a}\Delta q_{a}+\gamma_{b}\Delta q_{b}).  
 \label{Hamiltonian:SCC-2} 
\end{eqnarray}
\end{subequations}

\subsubsection{Consistency of the second order term}
From Eq.~(\ref{eq.Secular}), one can get a relation  
\begin{eqnarray}
 \sum_i^{\rm occ} \langle \Psi_i|H_0|\Psi_i\rangle = \sum_i^{\rm occ} \varepsilon_i  
      - \sum_{a\mu b\nu} q_{a\mu} \gamma_{a\mu,b\nu}\Delta q_{b\nu}.       \label{eq.LagrangeM-2} 
\end{eqnarray}
Combining Eqs.~(\ref{eq.Hamiltonian1})$\sim$(\ref{eq.Hamiltonian3}) with Eq.~(\ref{eq.LagrangeM-2}),
the total energy for the charge density $n_0+\delta n$ is obtained as  
\begin{subequations}
\begin{eqnarray}
&& 
E^{\rm LDA}[n_0+\delta n] \nonumber \\
&& \ \ \ \ \  = \Big[\sum_i^{\rm occ} \langle \Psi_i|H_0 |\Psi_i\rangle +E_{\rm onsite}+E_{\rm lattice}\Big] \nonumber \\
&& \ \ \ \ \  + \frac{1}{2}\sum_{a\mu b\nu} \Delta q_{a\mu} \gamma_{a\mu, b\nu} \Delta q_{b\nu} \label{eq.12_SCC-1} \\
&& \ \ \ \ \  = \Big[\sum_i^{\rm occ} \varepsilon_i + E_{\rm onsite}+E_{\rm lattice}\Big]  \nonumber \\
&& \ \ \ \ \ - {\frac{1}{2}\sum_{a\mu b\nu} (q_{a\mu} q_{b\nu}-q_{a\mu}^{(0)}q_{b\nu}^{(0)}) \gamma_{a\mu,b\nu} }  . \label{eq.12_SCC-2} 
\end{eqnarray}
\end{subequations}
This expression constitutes the key SCC total energy used in the present paper. 
Note that $\varepsilon_i$ are the Lagrange multipliers 
and represent the eigenenergies for the Kohn-Sham states of the TB Hamiltonian 
$H_{a\mu,b\nu} =H_{a\mu,b\nu}^{\rm 0-TB} +H_{a\mu,b\nu}^{\rm SCC}$.

\section{Three-parameter Formalism for Local Environment in TB-LDA}
Hereafter, $R_{ab}$ denotes the dimensionless ratio $R_{ab}/a_0$, where $a_0$ is the Bohr radius. 

\subsection{Definition and properties of local parameters $\varrho^a$, $r_s^a$, and $\eta^a$}
To represent the global and local environments, 
three parameters are introduced: 
a medium-range density ($\varrho^a$), 
an effective contact radius ($r_s^a$), 
and a local packing parameter ($\eta^a$).

The medium-range density $\varrho^a$ is defined as~\cite{NRL1,Fujiwara-Nishino-YamamotoECAL_2018}: 
\begin{subequations}
\begin{eqnarray}
 \varrho_a   &=& \sum_{b(\ne a)}e^{-\Lambda R_{ab}}F_c(R_{ab})  \label{varrho-1}  \\
 F_c(R_{ab}) &=& \frac{1}{1+\exp[(R_{ab}-R_0)/L]}           .      \label{varrho-2}
\end{eqnarray}
\end{subequations}
As $R_{ab}$ increases, the exponential term introduces a nonlinear dependence into this density parameter. 

An effective contact radius $r_s^a$ is defined as \cite{Fujiwara-Nishino-YamamotoECAL_2018}
\begin{eqnarray}
  r_s^a(m)
    \equiv \bigg[\sum_{b (\ne a)} (R_{ab})^{-m} \bigg]^{-\frac{1}{m}} .    \label{def-rsa-1}
\end{eqnarray}

In addition to $r_s^a$, a third parameter, referred to as the local packing parameter, is defined as
\begin{eqnarray}
&& \eta^a 
 = \frac{\partial \log r_s^a}{\partial \log m}  
 = -\log{r_s^a} +\sum_{b (\ne a)}p_{ab}(m) \log R_{ab}, \ \ \ \ \ \  \label{def-eta2} \\
&& p_{ab}(m)={(R_{ab})^{-m}}/{\sum_{c (\ne a)} (R_{ac})^{-m}}. \nonumber
\end{eqnarray}
Equation~(\ref{def-eta2}) is used in actual calculation, either crystals, random systems or 
in molecular dynamics (MD) simulation.

Let $R_{\rm nn}^a$ and $N_{\rm nn}^a$ denote the distance to and the number of nearest-neighbor (n.n.) atoms 
of atom $a$ in a crystal, respectively.
For clarity, Eq.~(\ref{def-eta2}) can be rewritten into the following explicit form, 
using nearest-neighbor quantities, although the definition itself is not restricted to crystalline systems: 
\begin{eqnarray}
 \eta^a 
 &=& \frac{1}{m}\log\Big[N_{\rm nn}^a\big\{1+\frac{1}{N_{\rm nn}^a}\sum_{b (\ne {\rm n.n.})}\Big(\frac{R_{\rm nn}^a}{R_{ab}}\Big)^m \big\}\Big]  \nonumber \\
 & & + \frac{1}{\sum_{c (\ne a)} \big(\frac{R_{\rm nn}^a}{R_{ac}}\big)^{m}}\sum_{b (\ne a)} \Big(\frac{R_{\rm nn}^a}{R_{ab}}\Big)^{m}\log \frac{R_{ab}}{R_{\rm nn}^a} . \ \ \ \ \ \  \label{def-eta3}     
\end{eqnarray}
Therefore, the local packing parameter $\eta^a$ characterizes geometrical coordination.

Furthermore, from Eq.~(\ref{def-eta3}), 
the following expression is derived for large $m$ or 
for the case in which the nearest-neighbor shell is isolated from further neighbor shells:
\[
  \eta^a \simeq (1/m)\log N_{\rm nn}^a 
\] 
The local packing parameter $\eta^a$ remains invariant under scaling $R_{ab}\rightarrow \kappa R_{ab}$, 
and this invariance ensures that $\eta^a$ characterizes geometry rather than length scale. 
Furthermore, the value of parameter $\eta^a$ changes sensitively 
with changing the environment of the near neighbor atoms. 
For a diamond structure, when $m \ge 15$, the contribution from second-neighbor atoms to the summation in Eq. (19) 
is of the order of $10^{-5}$. 
This consideration motivates the choice of $m = 20$ in the present study. 
Actual values of $\varrho^a$ and $r_s^a$ (or $1/r_s^a$) are listed in Table \ref{table:structural change}.

\subsection{Local environment parameters as generalized displacement coordinates}

The two parameters $\varrho^a$ and $r_s^a$ are linearly independent but correlated. 
They are sensitive to both the medium-range density and the local atomic environment, 
and together they provide a complementary description of structure-dependent physical properties.

\begin{figure}[ht]
\center{
\includegraphics[clip,width=7.5cm]{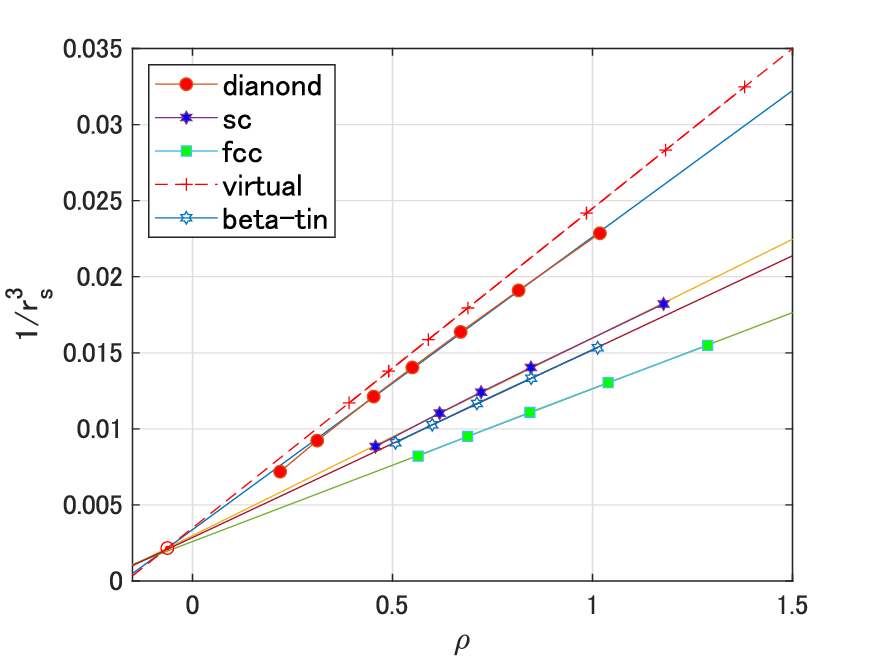}
}
\caption{(Color online) Phase diagram of diamond, fcc, sc, $\beta$-Sn, and VS as a function of $r_s^a$ and $\varrho^a$. 
$\Lambda=0.6$, $R_0=14.0$, $L=0.5$ in Eq. (\ref{varrho-1}) and $m=20$ in Eq. (\ref{def-rsa-1}). 
Sample points of the VS are not referred to fit band structures. 
}\label{Fig.StructureRho-Rs}
\end{figure}

Figure \ref{Fig.StructureRho-Rs} shows sample distributions in the $\varrho$-$1/{r_s}^3$ space 
for several silicon structures, 
including diamond structures, simple cubic (sc), face-centered cubic (fcc), and $\beta$-tin phases. 
In the figure, points corresponding to the virtual structure (VS) have been added. 
The VS is simply provided as a set of values and is not determined by an explicit atomic structure. 
The VS sample points are not used as reference data in the band-structure fitting, 
but are used for fitting the on-site energies to hypothetical extrapolated values 
and for ensuring its smooth variation in a region of fewer nearest-neighbor atoms 
than in the diamond structure. 

\begin{table}[h]
\centering
\caption{
Structural change of $\varrho$ - $1/{r_s}^3$ space {\it i.e.} $1={r_s}^3(p_1 \varrho +p_2)$.  
The intersection point $(\varrho_0, ({r_s}^{-3})_0)$ is given as 
$ \varrho_0=-0.0628, \  ({r_s}^{-3})_0=0.0022$.  $m=20$. 
}
\begin{tabular}{l|cccc} 
          & diamond   &  sc       &  fcc      &  $\beta$-tin \\
\hline  
  $p_1$     & 0.019239  & 0.012985  & 0.010031  &  0.012344    \\
  $p_2$     & 0.003375  & 0.002983  & 0.002595  &  0.002860    \\
 $\theta$   & 0.019293  & 0.012978  & 0.009799  &  0.012233    \\
 $\eta$     & 0.069409  & 0.090376  & 0.124460  &  0.088550    \\
 $\exp(m\eta)$ & 4.0075 & 6.0953    & 12.0516   &  5.8767      \\
\end{tabular}\label{table:structural change} 
\end{table}
The parameters used are $\Lambda=0.6$, $R_0=14.0$, $L=0.5$ for $\varrho^a$, and $m=20$ for $r_s^a(m)$. 
The data points of each structure lie approximately on a straight line 
\begin{eqnarray}
(1/{r_s}^3)=p_1 \varrho + p_2  . \label{varrho-rs relationship}
\end {eqnarray}
The three lines intersect (almost) at a single point $(\varrho_0, ({r_s}^{-3})_0)$. 
The corresponding parameter values are summarized in Table \ref{table:structural change}. 
A change along a straight line passing through this point 
corresponds to a homogeneous structural modification. 
In contrast, a transition from one line to another represents an inhomogeneous structural change, for example 
a change in crystal symmetry.
Thus, $\varrho^a$ and $r_s^a$ (or $1/r_s^a$) can thus be regarded as generalized displacement coordinates.

A uniform structural change can  also be described by two orthogonal coordinates $r$ and $\theta$;
\begin{subequations}
\begin{eqnarray}
  r &=& \sqrt{(\varrho-\varrho_0)^2+({r_s}^{-3}-({r_s}^{-3})_0)^2} , \label{eq.22a} \\
  \theta &=& \arctan\Big(\frac{{r_s}^{-3}-({r_s}^{-3})_0}{\varrho-\varrho_0}\Big). \label{eq.22b}
\end{eqnarray}
\end{subequations}
\begin{table*}[t]
\begin{minipage}{\textwidth}
\centering
\caption{
Band energy parameters of Si.
$\lambda=0.6$,  $R_0^{\rho}=14.0$,  $L^{\rho}=0.5$. 
$\varepsilon_{0}$ in eV unit and, $a$ and $b$ are in the atomic unit (a.u.).
}
\label{table:BandEnergy_parameter} 
{\scriptsize
\begin{tabular}{c|cccc} 
\hline
   $n,l$ & $\zeta_{(1)}$& $\zeta_{(2)}$&${C_{(1)}}/{C_{(2)}}$& $\varepsilon_{0}$~(eV)  \\
\hline 
   3s & 15.9617 & 1.7108  & 0.7288 & -19.0905  \\
   3p & 1.8575  & 1.3971  & 0.4510 & -11.9982  \\
   3d & 3.2557  & 1.1063  & 1.0899 & -7.1971   \\
\hline  
\end{tabular}
\begin{tabular}{c|cccccccc} 
\hline
   $n,l$ & $a$~(a.u.) & $b$~(a.u.) & $c$ & $d$ & $\exp(e)$& $\exp(g)$& $\exp(h)$& $\exp(k)$ \\
\hline 
   3s &   -0.5990 & 0.1604 & 0.3248 & -0.5446 & 0.0252  & 0.3128   & 0.5360   & 15.1087  \\
   3p &  -15.5005 & 0.3225 & 0.0156 & -0.5596 & 0.0919  & 0.3512   & 0.3236   & 1025.253 \\
   3d &   -4.7125 & 4.7299 & 0.0200 & -0.0394 & 6.42$\times 10^{-5}$ & 3.8266& 0.0925   & 0.5252 \\
\hline  
\end{tabular}
}
\end{minipage}
\end{table*}

\section{TB-matrix Elements and On-site Energy in $\varrho$, $r_s$, and $\eta$ Space}
Based on observations of the local density and packing parameters,
it is appropriate to use $\varrho^a$, $1/(r_s^a)^3$, and $\eta^a$
to describe both the band structure and the on-site energy. 

\subsection{TB-matrix elements $\Delta$ in $\varrho$ - $\eta$ space}
\subsubsection{Analytic expression and contour map}

At the equilibrium lattice constant of the diamond structure, $\Delta_\alpha(\varrho,\eta)$ is set to zero. 
Values of $\varepsilon_{0\alpha}$ are chosen to reproduce the band structure in ${\mib k}$ space, 
and are listed in Table \ref{table:BandEnergy_parameter}. 
The functional form of $\Delta_\alpha$ is expressed as a function of $\varrho$ and $\eta$: 
\begin{eqnarray}
 & & \Delta_\alpha(\varrho,\eta) = a \nonumber \\
 & & \  +b\frac{(\varrho+\exp(e))^c}{\Big[(\varrho+\exp(g))\{\tan(\eta)+\exp(h)\}+\exp(k)\Big]^d}, \ \ \ \ \ \ \ 
\label{eq.CorrectDelta}
\end{eqnarray}
where ($a, b, c, d, e, g, h, k$) are fitting parameters. 
Parameters are optimized to reproduce the band structures of silicon 
in the diamond, fcc, and sc phases by varying the lattice constants. 

Sample points need to cover multiple regions that cannot be fully represented by the diamond, sc, and fcc structures. 
Thus, three additional reference structures, derived from the diamond and fcc structures, 
are employed in fitting the band structure. \\
(1) An fcc structure with a large nearest-neighbor distance: \\
\ \ \ \ \ 
$
 \varrho = 0.0011,~~\eta = 0.1244,~~1/r_s^3 = 5.1381 \times 10^{-4}.
$ \\ 
(2) A distorted diamond structure, in which
the second fcc sublattice is shifted along the (111) direction: \\
\ \ \ \ \ 
$
 \varrho= 0.5598,~~\eta = 0.0025,~~1/r_s^3 = 0.0223. 
$ \\ 
(3) Anisotropic rectangular lattice: \\
\ \ \ \ \ 
$
 \varrho = 0.3301,~~\eta = 0.0346,~~1/r_s^3 = 0.02323.
$ \\ 
These additional sample points help to ensure system stability against local distortions. 
The optimized fitting parameters are summarized in Table~\ref{table:BandEnergy_parameter}.

\begin{figure}[h]
\center{
\includegraphics[clip,width=8.7cm]{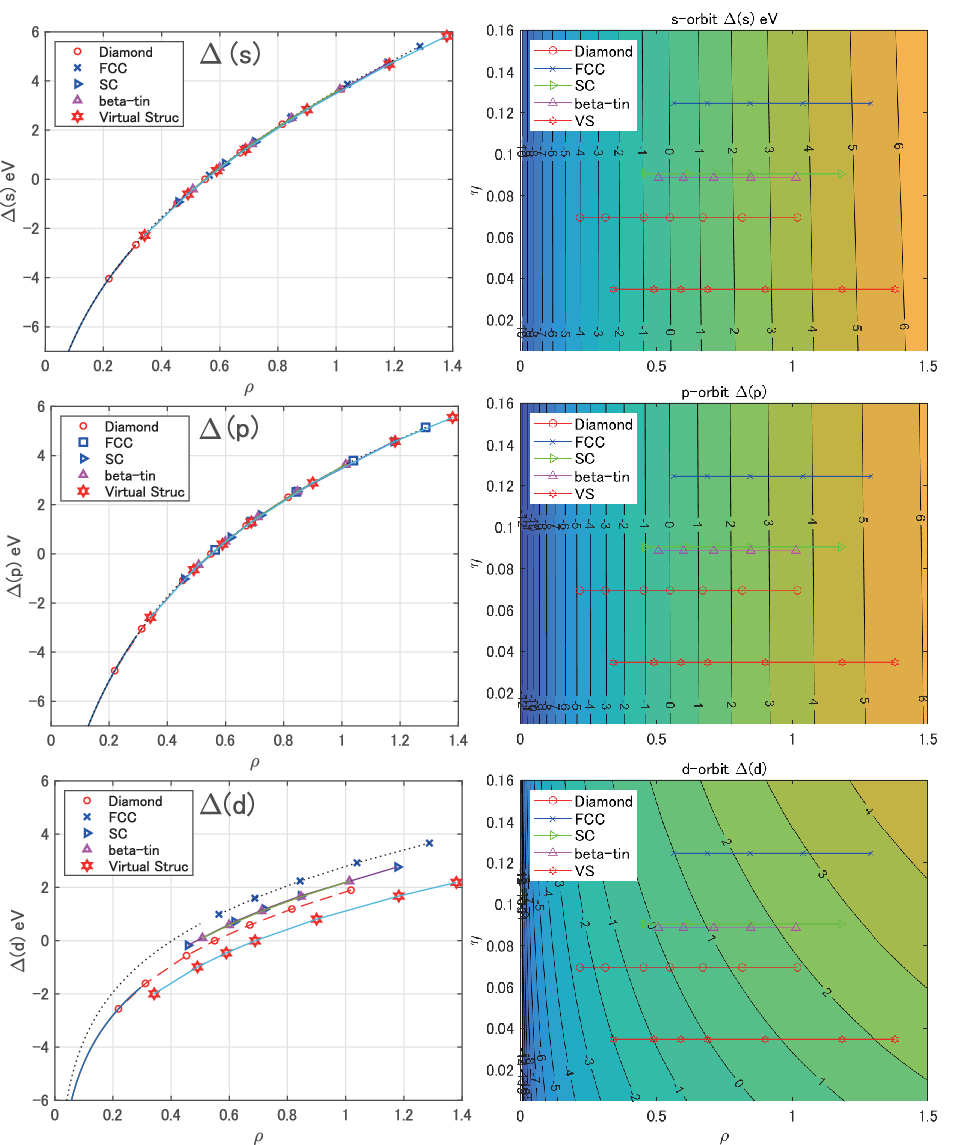}
}
\caption{(Color online) Variation of $\Delta$ (eV) (top;s, middle;p and bottom;d orbitals) 
as a function of $\varrho$ (left),  and the contour plot on $\varrho$-$\eta$ space (right).
Once the crystal structure is fixed, the values of the parameter $\eta$ remain constant 
while the density $\varrho$ varies (lines parallel to $\varrho$-axis). 
Sample points of the VS are not referred to get $\Delta$.
Note that $\delta \Delta$ is not added. }\label{Fig.Delta-spd}
\end{figure}

The dependence of $\Delta_\alpha(\varrho,\eta)$ is illustrated in Fig.~\ref{Fig.Delta-spd}. 
As shown, $\Delta_\alpha$ exhibits little or no dependence on $\eta$ in the s and p states, 
while a noticeable difference appears only in the d state, resulting in an almost uniform shift of the bands.

\subsubsection{Additional boundary condition for TB-matrix elements in the $\eta \sim 0$ region}
In the $\varrho$ - $\eta$ phase space, notably, 
the limit $\eta \rightarrow 0$ with a finite $\varrho^a$ corresponds to an isolated dimer. 
In this regime, the atomic levels should be higher than in the condensed phases. 
To ensure this expected behavior, a term $\delta \Delta$ is added to $\Delta(\varrho, \eta )$: 
\begin{eqnarray}
 \delta \Delta 
&=&
    \left\{
     \begin{array}{lll}
        1 & 0.0 < \eta \le 0.05  \\
        1-\Big\{1-\big(\frac{0.07-\eta}{0.02}\big)^2\Big\}^4 & 0.05 < \eta \le 0.07    \\
        0 & 0.07 < \eta        \\
   \end{array} \right. \nonumber  \\
& & \times 0.780,  \ \  ({\rm in \ \ eV}). \label{delta-Delta}
\end{eqnarray}
The correction (\ref{delta-Delta}) becomes significant when the effective number of the nearest-neighbor atoms falls below 3 
($\eta< 0.06$). 
Therefore, $\delta\Delta$ is important in stabilizing surface structures 
by preventing atoms from dissociating from surfaces. 

\subsubsection{Band energy as a function of n.n. distances}

\begin{figure}[h]
\center{
\includegraphics[clip,width=7.5cm]{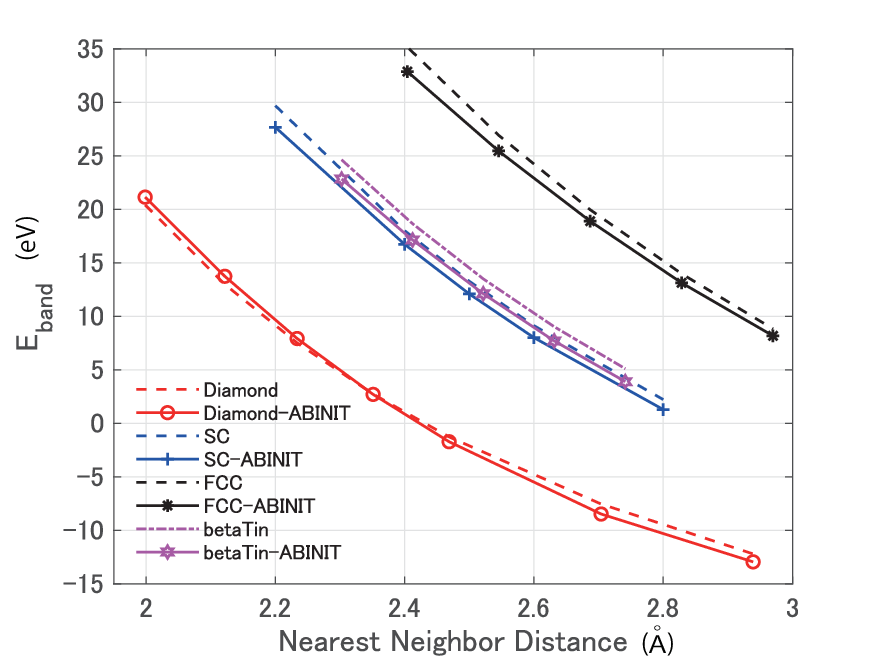}
}
\caption{(Color online) Band energy $E_{\rm band}$ (eV) 
as a function of nearest-neighbor distances (\AA).
}\label{Fig.Eband}
\end{figure}

After applying a uniform 20 eV offset to all bands of all crystal structures~\cite{J_Cerda_01}, 
the resulting band structures are in good agreement with LDA results, 
as reported in our previous work.~\cite{Fujiwara-Nishino-YamamotoECAL_2018}
The present formulation achieves a similar level of accuracy.
For brevity, the detailed band structures are not shown here; 
instead, Fig.~\ref{Fig.Eband} compares the band energies obtained from the present method 
with those from the ABINIT calculations. 
\subsection{On-site energy in $\varrho$ - ${1}/{{r_s}^3}$ space}
The on-site energy is constructed under the same design principles as the band energy, 
and is chosen as a smooth function of $(\varrho, 1/r_s^3)$ to avoid being overly influenced by local packing structure. 
\subsubsection{Interpolation of on-site energy}
The on-site energy at one atom site is assumed to be expressed as 
\begin{eqnarray}
&& E_{\rm onsite}^a(x,\theta)={\cal Q}_1(\theta)+{\cal Q}_2(\theta)(sx)+ {\cal Q}_3(\theta)(sx)^2  \nonumber \\ 
&& + {\cal Q}_4(\theta)h_3(x) + {\cal Q}_5(\theta)h_4(x) + {\cal Q}_6(\theta)h_5(x)      \nonumber \\ 
&& + {\cal Q}_7(\theta)h_6(x) + Q_8h_7(x) ,    \ \ \ Q_8=1        \label{def-EonsiteOnr} 
\end{eqnarray}
The constant ${\cal Q}_1$ is common across all structures. 
The other coefficients ${\cal Q}_n(\theta)$ ($n=2 \sim 7$) 
are optimized individually for each fixed crystal structure ({\it i.e.} fixed $\theta$). 
\subsubsection{Analytic expression of $E_{\rm onsite}$}
\begin{figure}[bht]
\center{
\includegraphics[clip,width=8cm]{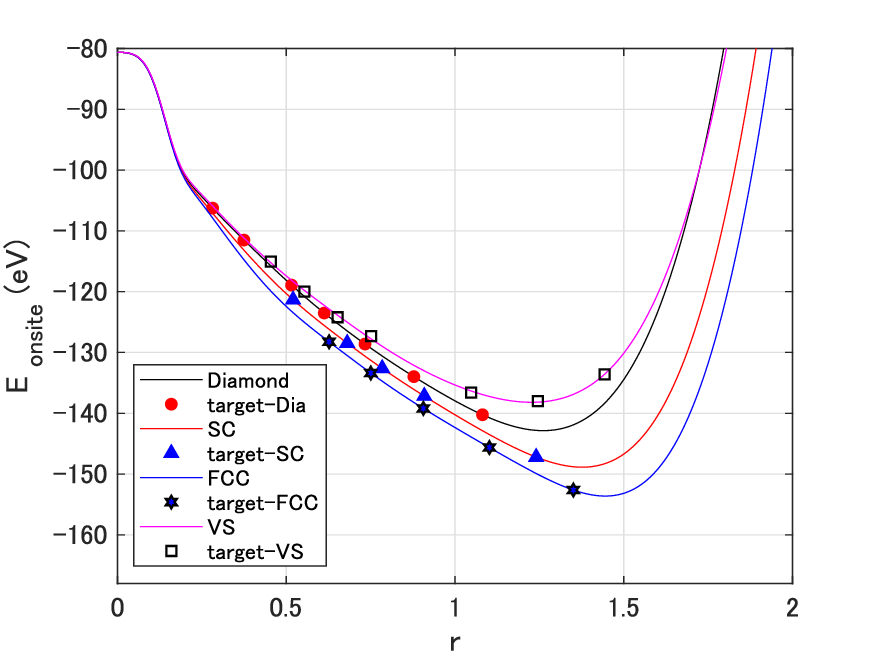}
}
\caption{(Color online) $E_{\rm onsite}$ in Eq. (\ref{def-EonsiteOnr}) for several reference systems (constant $\theta$). 
Marks indicate the target values. 
$r=\sqrt{(\varrho-\varrho_0)^2+(1/{r_s}^3-(1/{r_s}^3)_0)^2}$. }\label{Fig2P.StructureRho-Rs}
\end{figure}

\noindent
{{\bf Functions $h_n(r^2)$ in fixed crystal structures : }} 
The essential features of the on-site energy $E_{\rm onsite}$ for each crystal structure
are first approximated by a linear combination of $1,~x~(\equiv r^2)$, and $x^2$, 
and further refined using auxiliary functions $h_n(x)$ ($n=3 \sim 7$):
\begin{subequations}
\begin{eqnarray}
          h_n(x) &=& (sx)^3\exp\Big\{ h_{n2}\big(sx+h_{n1}\big)^2-h_{n1}^2 h_{n2} \Big\} ,   \nonumber \\
                 & & \ \ \ \ \ \ \ \ n=3 \sim 6 \ \ ,                                    \label{def-hn(r)} \\
          h_7(x) &=& h_{70}\exp(h_{72}(sx)^2).                                              \label{def-h7}
\end{eqnarray}
\end{subequations}
A scale factor $s$ is employed so that $h_n(x)$ can serve as universal functions and is applicable to a wide range of materials. 
In the present work, $s=2.0$ is used for condensed silicon. 
The resulting parameters for $h_3 \sim h_6$ are summarized in Table \ref{table:BaseFunc_parameter}. 
Actual feature of $E_{\rm onsite}$ for several reference systems (constant $\theta$) is depicted in Fig.~\ref{Fig2P.StructureRho-Rs}.

\begin{table}[b]
\centering
\caption{
Parameters for base functions of crystalline Si.
}
\label{table:BaseFunc_parameter} 
\center{
\begin{tabular}{c|ccccc} 
\hline
            & $h_3(x)$ & $h_4(x)$ & $h_5(x)$ & $h_6(x)$ & $h_7(x)$ \\
\hline 
   $h_{n0}$ & -        & -        & -        & -        &  16.5444 \\
   $x_n$    & 0.1964   & 0.8732   & 1.8062   & 0.09     &  -       \\
   $h_{n1}$ & 4.5758   & 9.8628   & 5.1144   & 8.2433   &  -       \\
   $h_{n2}$ & -1.6     & -0.16    & -0.12    & -2       & -2000    \\
\hline  
\end{tabular}
}
\end{table}

\noindent
{{\bf Coefficient $Q_m(\theta_k)$ at reference systems : }} 
The coefficient $Q_1(\theta)$ is set constant, and 
all coefficients ${\cal Q}_m$ ($m=1 \sim 7)$ are determined for reference systems of fixed $\theta_k$'s, 
using the Levenberg-Marquardt algorithm. 
The reference systems are fcc, sc and diamond structure for Si and their $\theta$ are 
$\theta_{\rm fcc}=0.098$, $\theta_{\rm sc}=0.0130$, and $\theta_{\rm dia}=0.0193$.

\noindent
{{\bf Akima spline interpolation for $Q_m(\theta_k)$ : }} 
The $\theta$-dependence of ${\cal Q}_m(\theta)$ is interpolated by the Akima algorithm~\cite{Akima-1-1970,Akima-2-1974}, 
which provides piecewise polynomials with continuous first and second derivatives. 
This algorithm is numerically stable and well-suited for handling oscillatory data 
as well as data with rapidly varying second derivatives. 
The resulting piecewise polynomials are also used in evaluating atomic forces analytically. 

\subsubsection{Additional boundary conditions for $h_n(x)$ and $Q_m(\theta$)}
{\bf Additional boundary condition for $h_n(x)$ for $x\rightarrow 0$ and $x\rightarrow \infty$}: 
The term $h_7(x)$ adjusts $E_{\rm onsite}$ to match the reference-extrapolation limit $r \rightarrow 0$. 
The value $E_{\rm onsite}(r\simeq 0)$ is evaluated as $-80.62$~eV from ABINIT calculation at a large lattice constant. 
Consequently, ${\cal Q}_1=-97.1644$~eV, $h_{70}=(E_{\rm onsite}(r\simeq 0)-{\cal Q}_1)=16.5444~{\rm eV}$, and $h_{72}=-2000$. 
The boundary condition for the opposite limit $r \rightarrow \infty$ is naturally described by the $x^2$ term

{\bf Additional boundary condition for $Q_n(\theta)$ in the regions $\theta \sim 0$ and 
$\theta > 0.3$: }
An additional stability condition may be imposed as $E_{\rm onsite}$ in the regions 
$\theta \rightarrow 0$ and $\theta \rightarrow \pi/2$. 
Possible instability in force field (at a singularity $r=0$) is outside of physical region. 
Our trial does not sample local structure in such region. 

\begin{figure}[t]
\center{
\includegraphics[clip,width=8cm]{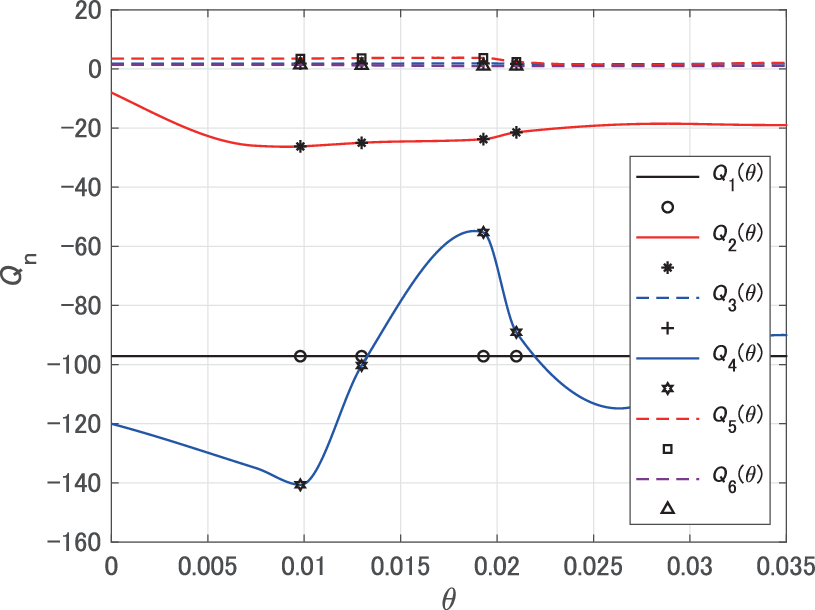}
}
\caption{(Color online) Inter- and extrapolated coefficients ${\cal Q}_1(\theta) \sim {\cal Q}_6(\theta) $. 
The marks indicate the target values of ${\cal Q}_n(\theta^{(k)})$\ ($k=3 \sim 6$). 
{${\cal Q}_n$ are in eV unit. 
$Q_1(\theta)$ is set constant over whole region. }
}\label{Fig2PP.StructureRho-Rs}
\end{figure}

In practice, seven reference points of $\theta$ are employed: 
$\theta^{(1)} =0$, 
$\theta^{(2)}=0.0075$, 
$\theta^{(3)}=\theta_{\rm fcc}=0.098$, 
$\theta^{(4)}=\theta_{\rm sc}=0.0130$, 
$\theta^{(5)}=\theta_{\rm dia}=0.0193$ , 
$\theta^{(6)}=0.0210$, and 
$\theta^{(7)}=0.035$. 
The $\theta^{(6)}$, referred to as the `virtual structure' (VS), 
represents a point just outside the crystal (i.e., in the vacuum region when the system contains a surface). 
The $\theta$-dependence of ${\cal Q}_n(\theta)$ is illustrated in Fig.~\ref{Fig2PP.StructureRho-Rs}.
The residual error of $E_{\rm onsite}$ at each $\theta$ point is well controlled. 

\subsubsection{Contour map of $E_{\rm onsite}(r, \theta)$}

\begin{figure}[h] 
\center{
\includegraphics[clip,width=9.0cm]{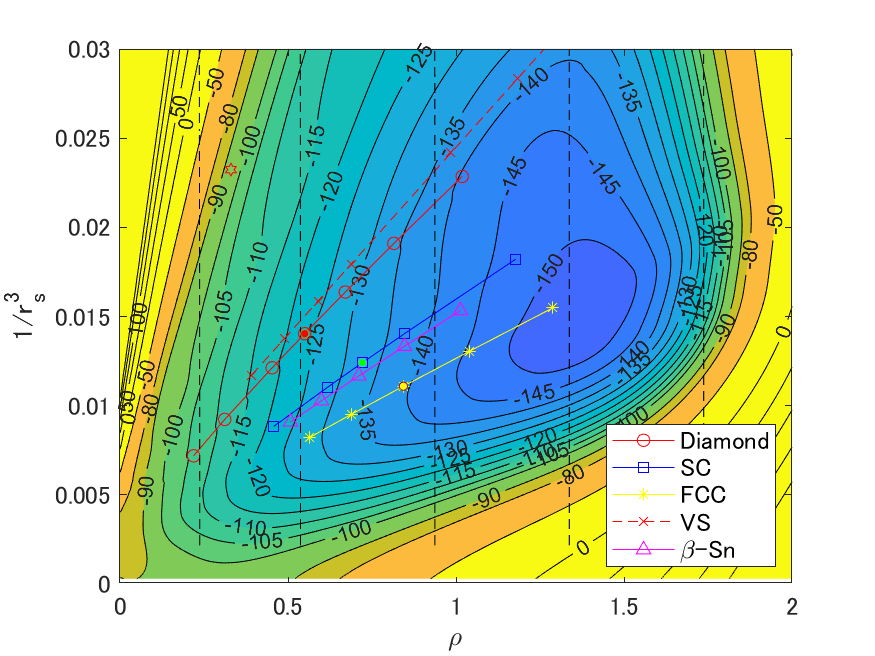}
\includegraphics[clip,width=6.5cm]{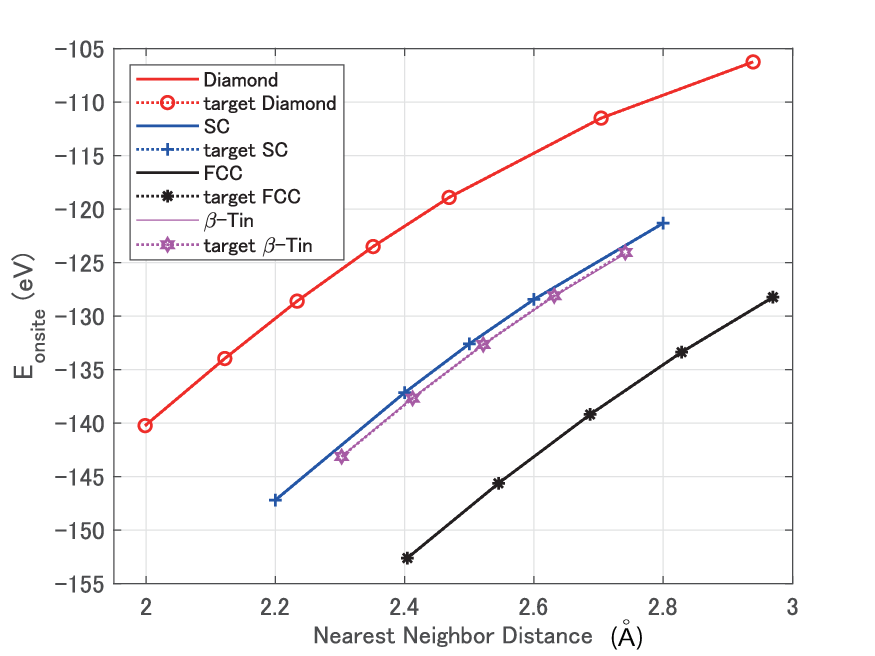}
}
\caption{(Color online) 
(a) $E_{\rm onsite}$ as a function of $\varrho$ and $r_s$.
The contour plot of $E_{\rm onsite}$ is shown in eV units. 
Marks indicate the positions used as reference.
The closed points show the positions of the minimum total energy in each structure.
The chain curves indicate contours of constant $r$. ($r=0.3, 0.6, 1.0, 1.4, 1.8$). 
Note that, as the scales of the abscissa and the ordinate are different, 
the curves of constant $r$ do not appear to be perpendicular to the line for each crystal (curve of constant $\theta$).
(b)
$E_{\rm onsite}$ (eV) as a function of the nearest-neighbor distance (\AA). 
The marks indicate the values of the ABINIT calculation.
}
\label{Eonsite-rho-rs-2}
\end{figure}

\begin{figure}[h]
\center{
\includegraphics[clip,width=6.5cm]{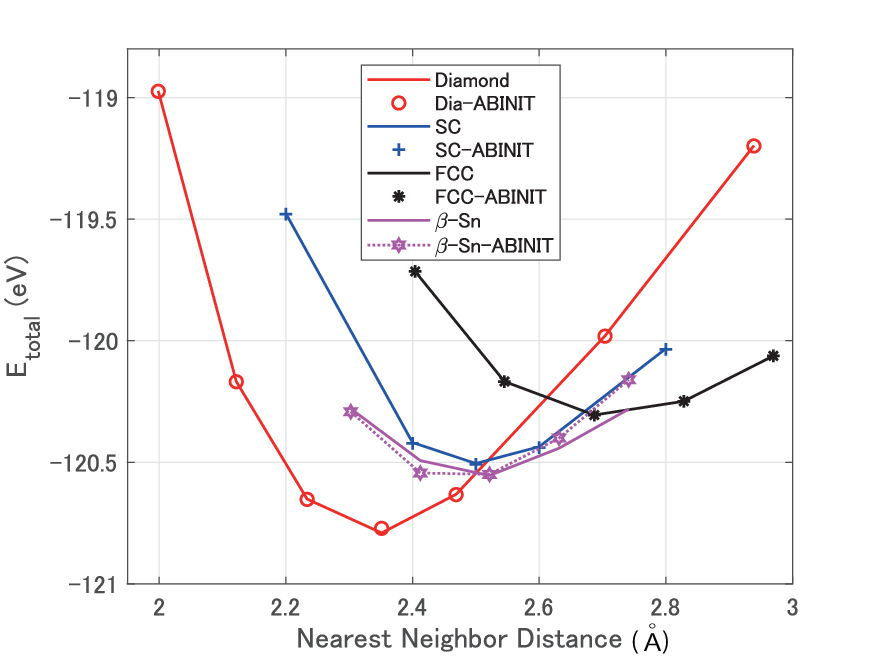}
}
\caption{(Color online) 
$E_{\rm total}$ (eV) as a function of the nearest-neighbor distance (\AA). 
The marks indicate values obtained from the ABINIT calculations.
}
\label{E_onsiteE_total-2}
\end{figure}

Figure \ref{Eonsite-rho-rs-2} shows the calculated on-site energy $E_{\rm onsite}(r, \theta)$, 
obtained from Eq.~(\ref{Def-trueEonsite}) in $\varrho$ - $1/{r_s}^3$ space. 
The overall behavior of the on-site energy 
$E_{\rm onsite}$ varies only slowly and slightly for $\theta_{\rm fcc}<\theta < \theta^{(6)}$ 
when $r$ is kept constant within the range $r=0.2 \sim 0.7$. 
This result suggests that changes in the local structure (i.e., variations in $\theta$) 
are primarily governed not by the on-site energy, 
but rather by fluctuations in the band-structure energy. 
The results of $E_{\rm onsite}$ and $E_{\rm total}$ are presented in Fig.~\ref{E_onsiteE_total-2}.
These results were obtained after optimizing the parameters 
using the Si diamond, fcc, and sc phases with VS hypothetical extrapolation data as references. 
\subsection{Total energy}
The discrepancy in $E_{\rm total}$ vanishes for all cases except the $\beta$-tin structure. 
This small deviation does not arise from the exclusion of the $\beta$-tin structure as a reference system,
but rather due to its high anisotropy. 
Indeed, when the $\beta$-tin structure was included as a reference, 
a small anomaly (or spike) appeared in all ${\cal Q}_n(\theta)$ at $\theta=0.01223$, 
corresponding to the $\beta$-tin structure. 

\subsection{Forces}
The force acting on an atom $a$ can be decomposed as
\begin{eqnarray}
 F^a &=& -\frac{\partial E_{\rm total}}{\partial R_a} \nonumber \\
     &=& F^a_{\rm band}  +F^a_{\rm SCC}+ F^a_{\rm onsite} +F^a_{\rm lattice} . \label{eq.force1}
\end{eqnarray}
Each term in Eq.~(\ref{eq.force1}) comes from 
$H^{\rm 0-TB}$, $H^{\rm SCC}$, $E_{\rm onsite}$, and $E_{\rm lattice}$, respectively. 
Explicit expressions of $F^a_{\rm band}$ and $F^a_{\rm SCC}$ can be found in Ref.~\cite{M_Elstner_01,Witek2004} and in Appendix. 
The forces are also explicitly written down as analytic functions of $\varrho^a$, $1/(r_s^a)^3$, and $\eta^a$ in Appendix. 

\subsection{Actual workflow for determination of local parameters}
All TE-TB parameters are determined so as to reproduce the band structures
and total energies for as many crystalline structures as possible,
including both existing and hypothetical crystalline structures. 
In our tests, we did not encounter instabilities attributable to the choice of initial values.

\begin{description}
\item[Step 0]
   Choose the principal structure S$_0$ (e.g., the diamond structure in the case of Si). 
\item[Step 1]
   For the principal structure S$_0$ with a fixed lattice constant, 
   determine the values of the STO parameters
    ($\zeta_1$, $\zeta_2$, $C_{(1)}/C_{(2)}$, $\varepsilon_0$)  
   so as to obtain good agreement between the first-principles (FP) calculations and the present TE-TB calculations 
   for the $E$-${\mib k}$ relation $E_n({\mib k})$ and the band energy. 
\item[Step 2]
   Then determine all parameters in $\Delta(\varrho, \eta)$ 
   to obtain good agreement of 
   $E$-${\mib k}$ relation and the band energy $E_{\rm band}^{\rm TE-TB}|$ for varying lattice constants and 
   crystal structures S$_0$, S$_1$, S$_2$, S$_3$ $\cdots$.
   \begin{eqnarray*}
     E_n({\mib k})^{\rm TE-TB}|_{{\rm S}_i} & \simeq & E_n({\mib k})^{\rm FP}|_{{\rm S}_i} \\
     E_{\rm band}^{\rm TE-TB}|_{{\rm S}_i}  & \simeq & E_{\rm band}^{\rm FP}|_{{\rm S}_i} \ \ \ i=0, 1, 2,3,\cdots \ \ .
   \end{eqnarray*}
\item[Step 3]
   Calculate $E_{\rm lattice}^{\rm TE-TB}$ in Eq.~(\ref{Lattice-Coulomb}) for structures S$_0$, S$_1$, S$_2$, S$_3$ $\cdots$ 
   using the resulting electron charge density.  
\item[Step 4]
   Using the calculated $E_{\rm band}^{\rm TE-TB}$ and $E_{\rm lattice}^{\rm TE-TB}$, 
   evaluate the on-site term in the TE-TB method as
   \begin{eqnarray*}
     E_{\rm onsite}^{\rm TE-TB}\Big|_{{\rm S}_i} = 
        \{E_{\rm total}^{\rm FP} - E_{\rm band}^{\rm TE-TB} -E_{\rm lattice}^{\rm TE-TB}\}\Big|_{{\rm S}_i}\ .  \label{Step5}
   \end{eqnarray*}
   Then, a set of local energy parameters should be determined   
   in the $\varrho$-$1/r_s^3$ space to satisfy 
   \begin{equation*}
     E_{\rm onsite}(r_s,\varrho)\Big|_{{\rm S}_i} =  E_{\rm onsite}^{\rm TE-TB}\Big|_{{\rm S}_i}  \ \ (i=0,1,2 \cdots) .
   \end{equation*}
\item[Step 5]
   Calculate
   \begin{eqnarray*}
     E_{\rm total}^{\rm TE-TB} \Big|_{{\rm S}_i}\equiv
         \big\{E_{\rm band}^{\rm TE-TB} + E_{\rm lattice}^{\rm TE-TB} +E_{\rm onsite}(r_s,\varrho) \big\}\Big|_{{\rm S}_i}   \label{TE-TB_TotalE}
   \end{eqnarray*}
   for S$_0$, S$_1$, S$_2$, S$_3$ $\cdots$. 
   Because of the definition in Step 4, good agreement for $E_{\rm total}$ is expected;
   \begin{eqnarray*}
      E_{\rm total}^{\rm TE-TB}\Big|_{{\rm S}_i} & \simeq & E_{\rm total}^{\rm FP}\Big|_{{\rm S}_i}\ \ \ i=0, 1, 2, 3,\cdots .
   \end{eqnarray*}
\item[Step 6]
    SCC correction should be incorporated in or after the parameter-determination  procedure.
    In the present work of Si, the SCC correction is not necessary in the process of parameter-determination. 
\end{description}
 After all parameters are determined in bulk systems and, in Sec.~\ref{Demonstration}, the SCC term is introduced 
 in calculations of monovacancy and surface.

\section{Demonstration for Crystalline Silicon}\label{Demonstration}
The present formalism is applied to three representative cases in diamond-structured silicon: 
(1) the stability of the diamond structure, 
(2) the formation of a monovacancy, and 
(3) the reconstruction of the (001) surface. 

The on-site Coulomb parameter $\gamma_{a}=U_a$ 
is estimated from the difference between the ionization energy and electron affinity, 
giving $U_a=9.2~{\rm eV}$.
Equations (\ref{eq.12_SCC-1}) and (\ref{eq.12_SCC-2}) require careful treatment of the off-diagonal SCC Coulomb term. 
For large interatomic separations $R_{ab}$, $\gamma_{a,b}$ becomes long-ranged and approaches $1/R_{ab}$.

When calculating the formation energy of a monovacancy using a supercell, 
the Coulomb interactions between periodically repeated vacancies must be excluded.
In other words, ion pairs whose interaction is dominated by other vacancies are omitted. 
Similarly, in a finite-thickness slab, 
the interactions between the upper and lower surfaces should be excluded.

The calculation proceeds in two steps. \\
First, Eq.~(\ref{eq.12_SCC-1}) or (\ref{eq.12_SCC-2}) is evaluated under the simplified assumption 
$\gamma_{a\mu,b\nu}=\gamma_{a,a}\delta_{a,b}$. 
Atoms are moved according to the calculated force. 
The total energy is then calculated as 
\begin{subequations}
\begin{equation} 
   \Delta E_{\rm total}^{(1)}=E_{\rm total}[n_0+\delta n]\Big|_{\gamma_{a,b}=\gamma_{a,a}\delta_{a,b}}-E_{\rm total}^{\rm (bulk)}, \label{eq.dETotal1}
\end{equation}
$E_{\rm total}^{\rm (bulk)}$ is the total energy of the corresponding bulk system. 
\\
Second, the off-diagonal SCC Coulomb energy is added: 
\begin{equation}
  \Delta E_{\rm total}[n_0+\delta n] = \Delta E_{\rm total}^{(1)}+\sum_{\{a\ne b\}}\Delta q_a \gamma_{a,b}\Delta q_b . 
\label{eq.dETotal2}
\end{equation}
\end{subequations} 
In periodic vacancy models and slab surface models, 
great care must be taken with the off-diagonal SCC Coulomb energy 
to avoid artificial phenomena arising from boundary conditions. 
Furthermore, it should be noted that a charged defect 
(whether real or artificial, arising from the iterative calculation process) 
may locally compromise the positive definiteness of the SCC interaction matrix 
in the sense of the Gershgorin disc theorem. 
This may lead to local numerical instability in the Hamiltonian.
\subsection{Global stability}
The first test is straightforward: 
we confirm that the system rapidly returns to equilibrium after relaxation.
The SCC correction vanishes when all Si atoms are neutral. 

The procedure is as follows:
\begin{enumerate}
\item Construct a cubic diamond supercell of 512 Si atoms under periodic boundary conditions, 
with cell dimensions 21.72~\AA$\times$21.72~\AA$\times$21.72~\AA. 
\item Displace one sublattice of Si atoms along:  
(a) the direction toward a vacant T-site with T$_{\rm d}$ symmetry, 
(b) the opposite direction of case (a), and 
(c) the direction toward a vacant H-site with D$_{\rm 3d}$ symmetry.~\cite{Impurity-1999} 
\item Relax all Si atoms along the calculated force directions. 
\item Iterate until the total energy converges within $10^{-9}$~eV/atom. 
\end{enumerate}
All atoms return to their ideal positions, 
confirming that $E_{\rm total}[n]$ remains in equilibrium under the present formalism. 

\subsection{Defect and surface}
Further tests focus on defects and surfaces of the diamond structure. 
These tests may reveal the capabilities of the TE-TB method, 
although it is not specifically designed for detailed studies of defects or surface structures.

\subsubsection{Monovacancy}
This test examines whether a monovacancy structure can be properly relaxed.~\cite{Vacancy-1993, Vacancy-2015}

The steps are:
\begin{enumerate}
\item First, construct a cubic supercell consisting of 511 Si atoms and one vacancy under periodic boundary conditions, 
with dimensions 21.72~\AA$\times$21.72~\AA$\times$21.72~\AA. 
\item Next, relax all Si atoms along the force directions. 
\item Repeat this process until the total energy converges within $10^{-6}$~eV/atom. 
\item Finally, vacant region is analyzed 
with introducing the off-diagonal SCC Coulomb term $\sum_{\{a\ne b\}}\Delta q_a\gamma_{a,b}\Delta q_b$.
\end{enumerate}
After the above steps are completed, the vacant region contracts while preserving {\bf Td symmetry}. 
The distance between the vacancy center and its nearest-neighbor Si atoms is 1.58~\AA, 
while the edge length of the resulting tetrahedron is 2.58~\AA. 
The Mulliken charge of an atom nearest to the vacancy was calculated to be 3.91. 

With a chemical potential of the system of 512 atoms $\mu=E_{\rm total}^{\rm (bulk)}(512~{\rm atoms})/512$, 
the formation energy, neglecting the off-diagonal SCC energy (up to Step (3)), is evaluated as 
\begin{eqnarray}
  && E_{\rm total}(511~{\rm atoms})-E_{\rm total}^{\rm (bulk)}(512~{\rm atoms})+\mu \nonumber \\
  && =E_{\rm total}(511~{\rm atoms})-\frac{511}{512}\times E_{\rm total}^{\rm (bulk)}(512~{\rm atoms}) \nonumber \\
  && \simeq 0.0676 \  ({\rm eV}). 
\end{eqnarray}
At Step (4), adding the \ off-diagonal SCC energy 1.1646~eV, 
the formation energy is estimated as 1.2322~eV.

In order to study the neglected effects of the force due to the off-diagonal SCC energy, 
the remaining force on an atom $a$ is assumed to be approximated (see Appendix A1) as 
\begin{equation}
  {\mib F}^a_{\rm SCC-off} \simeq  - \Delta q_a \sum_c \frac{\partial \gamma_{a c}}{\partial {\mib R}_a}\Delta q_c . \label{eq.29b}
\end{equation}
Consider the effects of ${\mib F}^a_{\rm SCC-off}$  
on the four atoms in the first-shell and the twelve in the second-shell, surrounding the vacancy center. 
The force acting on each atom in the first shell is directed from the center to that atom. 
We evaluated the additional energy shift of the off-diagonal SCC contribution 
by choosing a displacement of $\Delta z=0.1$~\AA~ for the four first-shell atoms, 
with corresponding displacements for the other atoms. 
For reference, the distance an atom typically moves in a single step of  iteration 
is no more than $2\times 10^{-4}$ \AA.  
The work estimated from the residual force by the off-diagonal SCC energy was found to be 
\[
 \Delta E\simeq -0.038~{\rm eV/vacancy},
\] 
which is 3\% of the estimated formation energy.
In other words, the force contribution from the off-diagonal SCC term is very small. 
Therefore, it can be concluded that the effects of forces by the off-diagonal SCC energy 
to the estimated formation energy do not change the conclusion here. 
The estimated value of the formation energy is comparable to those obtained from LDA calculations.\cite{Vacancy-2015,Spiewak}

\subsubsection{{\rm (001)} surface}

\begin{table*}[ht]
\begin{minipage}{\textwidth}
\centering
\caption{Surface energies of dimerization in Si (001) surface. 
The first row $\Delta E_{\rm total}^{(1)}$ is the total energy measured from that of the crystalline state of the same size 
as defined in Eq.~(\ref{eq.dETotal1}). 
Others are explained in the text.
}
\begin{tabular}{l|rrrrrr} 
                                       & ideal  $p(1\times 1)$ & sym. $p(2\times 1)$ & asym. $p(2\times 1)$ & $p(2\times 2)$ & $c(4\times 2)$  \\
\hline  
 (1) $\Delta E_{\rm total}^{(1)}$/half slab (eV)& 285.8240 & 226.2660 & 145.7095 & 144.4591 & 144.1123  \\
 (2) $\Delta q \gamma \Delta q$/half slab (eV)  &   4.7450 &   1.5687 &  -0.2022 &  -0.2835 &  -2.5386  \\
 (3) $\Delta E_{\rm total}$/half slab (eV)      & 290.5690 & 227.8347 & 145.5073 & 144.1756 & 141.5737  \\
 (4) $\Delta E_{\rm total}$/dimer (eV)          &  18.1606 &  14.2397 &   9.0942 &   9.0110 &   8.8484  \\
 (5) Relative energy  (eV)                      &   9.3122 &   5.3913 &   0.2459 &   0.1626 &   0       \\ 
\end{tabular}\label{table:dimerE} 
\end{minipage}
\end{table*}

The most challenging example is the (001) surface of diamond-structured silicon. 
Ramstad and co-workers have investigated various dimer reconstructions on this surface, 
including the ideal $p(1\times 1)$, 
symmetric $p(2\times 1)$, 
asymmetric $p(2\times 1)$, $p(2\times 2)$, and $c(4\times 2)$.~\cite{(001)Surface-1995} 

In slab models involving surface problems, artificial effects arising 
from the interaction between the top and bottom surfaces can sometimes pose a serious problem. 
To address this issue, we initially omit the off-diagonal SCC terms and subsequently evaluate their effect.

We proceed as follows:
\begin{enumerate}
\item 
Construct a rectangular parallelepiped supercell consisting of 384 Si atoms 
under periodic boundary conditions in the a and b directions, with lattice vectors 
${\mib a}$=(21.72~\AA, 0, 0), ${\mib b}$=(0, 21.72~\AA, 0), ${\mib c}$=(0, 0, 42.89~\AA). 
The system consists of 12 (001) layers, each comprising 32 atoms in the $a$ - $b$ plane. 
\item 
Impose appropriate initial displacements of atoms on the upper and lower (001) surfaces 
corresponding to 
ideal $p(1\times 1)$, symmetric $p(2\times 1)$, asymmetric $p(2\times 1)$, $p(2\times 2)$ and $c(4\times 2)$ symmetry. 
\end{enumerate}
Once the above preparations have been completed, secure the central two layers in place, and 
follow the same relaxation protocol as for the monovacancy case. 
The total energy converges within $10^{-4}$~eV/atom

Relaxation results indicate that the ideal $p(1\times 1)$ structure is unstable. 
Its total energy remains metastable until up to about 800 iterations, 
and then decreases to an intermediate value between the metastable $p(1\times 1)$ 
and the stable symmetric $p(2\times 1)$ structures. 
The resulting atomic configuration is a mixture of these two structures. 
By contrast, the four reconstructed structures, 
symmetric $p(2\times 1)$, asymmetric $p(2\times 1)$,  $p(2\times 2)$, 
and $c(4\times 2)$, are stable. 

Table \ref{table:dimerE} lists the calculated energies. 
The first row (1) gives the values obtained 
by subtracting the total energy of a perfect six-layer crystal from that of a half-slab ({\it i.e.} total energy/2), 
considering only terms with $\gamma_{a,b}=\gamma_{a,a}\delta_{a,b}$.
The second row (2) reports the contribution from 
$\sum_{\{a\ne b\}}\Delta q_a \gamma_{a,b}\Delta q_b $ for pairs $\{a,b\}$, where
$a$ atoms on the surface layer (one among 32 atoms) and $b$ atoms on the half-slab. 
The third row (3) gives the sum of rows (1) and (2). 
The fourth row (4) presents the energies from row (3) divided by 16, {\it i.e.} the energies per surface dimer. 
The final row (5) shows the relative energies with respect to the lowest-energy structure. 

In the  $c(4\times 2)$ structure, 
the additional force arising from the off-diagonal SCC interaction 
acts on the surface atoms in the outward direction normal to the surface. 
The contribution of the off-diagonal SCC interaction to the potential energy 
associated with an additional displacement $\Delta z$ of a single top-layer atom 
along the direction of the additional force 
acting on that atom can be expressed as 
\begin{equation}
   \Delta E(\Delta z) = a\Delta z^2 + b\Delta z, \label{eq.Model}
\end{equation}
where 
$a= 1.3\times 10^{-3}~{\rm eV}/\AA^2$ and $b=-4.2\times 10^{-3}~{\rm eV}/\AA$. 
The small magnitudes of the coefficients $a$ and $b$ indicate that 
the contribution of the off-diagonal SCC interaction to the atomic relaxation is negligible.
For the $p(2\times 2)$ structure, 
the contribution of the off-diagonal SCC interaction is also small 
in the physically relevant displacement range.

For all the surface structures considered, 
the additional force arising from the off-diagonal SCC interaction is small.
Therefore, neither the energetic ordering nor the approximate energy differences 
among the different (001) surface structures are changed.

These results are consistent with those reported by Ramstad et al.~\cite{(001)Surface-1995}.

\section{Summary and Discussion}

The present study extends our previous work~\cite{Fujiwara-Nishino-YamamotoECAL_2018},
in which a tight-binding (TB) method was formulated on the basis of band structures
and total energies obtained within the local density approximation (LDA).
Here, the formalism is generalized by representing the total energy in terms of
three local parameters of $\varrho$, $1/r_s^3$, and $\eta$, together with additional boundary conditions. 
The framework has been verified through tests on bulk crystalline silicon,
a monovacancy in diamond-structured silicon, and the Si(001) surface.

Guidelines for extending DFT in the parameter space have been presented. 
Through physical considerations under explicitly imposed boundary conditions,
the effective transferability - interpreted here as smooth and controlled extrapolation 
in parameter space - has been significantly improved.
This design-oriented viewpoint may provide a useful basis for 
future methodological developments in materials physics.
The present TE-TB formalism is therefore expected to enable a controlled expansion
of the accessible parameter space, offering a practical foundation for future
materials exploration.

\appendix
\section{Calculation of Forces}

The explicit construction of force expressions from a constrained total-energy functional is essential 
for maintaining robustness under non-uniform and extrapolative structural variations.

In this Appendix, the explicit expressions for atomic forces are summarized. 
The present formalism is constructed at the level of the total energy functional, 
where physically motivated boundary conditions are explicitly imposed in the $\varrho^a$-$1/(r_s^a)^3$ space 
and the $\varrho^a$-$ \eta^a$ space. 
Therefore, the corresponding force expressions are uniquely and analytically determined by applying the chain rule, 
since all local environment parameters 
are explicit functions of atomic coordinates.
The resulting expressions can be implemented in the expressions of the standard SCC-TB force formalism~\cite{M_Elstner_01} 
and additional contributions. 
The atomic forces were explicitly evaluated over the entire parameter space, 
and every component was confirmed to form smooth vector field without numerical instabilities.

\subsection{Hamiltonian part $ {\mib F}^a_{\rm band}$ and SCC part ${\mib F}^a_{\rm SCC}$
}
The components of the force can be divided as shown in Eq.~(\ref{eq.force1}). 
The Hamiltonian part and SCC part are expressed as~\cite{M_Elstner_01}
\begin{eqnarray*}
 {\mib F}^a_{\rm band}
 &=& 
   -\sum_n^{\rm occ} \sum_{\mu}\sum_{b \nu} 
    \frac{1}{2}\big({c_{a \mu}^{(n)}}^* c_{b \nu}^{(n)}+{c_{b \nu}^{(n)}}^*c_{a \mu}^{(n)}\big) \nonumber \\
 && \ \ \ \ \times \Big\{\frac{\partial{(H_0)}_{a\mu,b \nu}}{\partial{\mib R}_a} 
         - \varepsilon_n \frac{\partial S_{a\mu,b \nu}}{\partial{\mib R}_a}   \Big\}  \label{0-th} \\
 {\mib F}^a_{\rm SCC}
 &=& 
   -\sum_n^{\rm occ}\sum_{b(\ne a)}
    \frac{1}{2}\big({c_{a \mu}^{(n)}}^* c_{b \nu}^{(n)}+{c_{b \nu}^{(n)}}^* c_{a \mu}^{(n)}\big) \nonumber \\
 && \ \ \ \ \times \frac{\{ \sum_c ( \gamma_{a,c}+\gamma_{b,c}) \Delta q_c\}}{2}
               \frac{\partial S_{a\mu,b\nu}}{\partial{\mib R}_a}                                 \nonumber \\
 && - \Delta q_a \sum_c
     \frac{\partial \gamma_{a,c}}{\partial {\mib R}_a}\Delta q_c .                  \label{2-nd} 
\end{eqnarray*}
These are the usual TB-force without charge fluctuation and the additional 2-nd order SCC term. 

\subsection{Local part ${\mib F}_{\rm onsite}$}
The force due to on-site (local) energy 
can be calculated by the expression 
\begin{eqnarray*}
 {\mib F}_{\rm onsite}^a  
  = -\Big\{ \frac{\partial E_{\rm onsite}^a}{\partial \varrho^a} \cdot \frac{\partial \varrho^a}{{\partial {\mib R}_a}}
    + \frac{\partial E_{\rm onsite}^a}{\partial 1/(r_s^a)^3}  \cdot \frac{\partial 1/(r_s^a)^3}{{\partial {\mib R}_a}}  \Big\} ,
\label{eq.force-Local}
\end{eqnarray*}
where $E_{\rm onsite}^a$ is the local on-site energy at the atom $a$.
The on-site energy $E_{\rm onsite}^a(x^a,\theta^a)$ is given as a local function of $x^a$ and $\varrho^a$ defined 
in Eqs.~(\ref{eq.22a}) and (\ref{eq.22b}).

Derivatives $\partial E_{\rm onsite}^a/\partial \varrho^a$ and $\partial E_{\rm onsite}^a/\partial (1/r_s^a)^3$ 
are given as 
\begin{eqnarray*}
&& \frac{\partial  E_{\rm onsite}^a(x^a,\theta^a)}{\partial \varrho^a} \\
&& =  2(\varrho^a-\varrho_0) \frac{\partial  E_{\rm onsite}^a}{\partial x^a}
    -\frac{\sin\theta^a \cos\theta^a}{\varrho^a-\varrho_0} \frac{\partial  E_{\rm onsite}^a}{\partial\theta^a}, \\
&& \frac{\partial  E_{\rm onsite}^a(x^a,\theta^a)}{\partial (1/r_s^a)^3} \\
&& =2\Big\{\frac{1}{(r_s^a)^3}-\Big(\frac{1}{r_s^3}\Big)_0\Big\}\frac{\partial  E_{\rm onsite}^a}{\partial x^a}
   + \frac{\cos^2\theta^a}{\varrho^a-\varrho_0} \frac{\partial E_{\rm onsite}^a}{\partial\theta^a}.
\end{eqnarray*}
It should be noted the singular point $\varrho^a=\varrho_0$ locates out of the physical region in an 
example in the text, since $\varrho_0<0$.

The actual on-site energy is a function of $x$ and $\theta$, as shown in Eq.~(\ref{def-EonsiteOnr}), as 
\begin{eqnarray*}
 E_{\rm onsite}^a (x, \theta) &=& \sum_{\{n,m\}} Q_n({\rm Akima \ polynomial\ of \ }\theta) \\
                               & &\times h_m({\rm function \ of} \ x). 
\end{eqnarray*}
The Akima polynomial is given as the third order one, and the function $h_m(x)$ is explicitly defined, so that 
the derivatives are directly written down. 

\subsection{Coulomb neutral part ${\mib F}_{\rm lattice}$}
The force due to Coulomb energy $E_{\rm lattice}$ is as follows;
\begin{eqnarray*}
&& {\mib F}_{\rm lattice}^a \\ 
&& = -\frac{\partial}{\partial {\mib R}_a}
   \sum_{\{b \ne c\}} 
   \Big\{\frac{Z_bZ_c}{|{\mib R}_b-{\mib R}_c|}
   -\int d{\mib r}\int d{\mib r}^\prime  \frac{n_b({\mib r})n_c({\mib r}^\prime)}{|{\mib r}-{\mib r}^\prime|} 
   \Big\}   .
\label{eq. force-Neutral}
\end{eqnarray*}
The second term is evaluated in the same way as in the previous work.~\cite{Fujiwara-Nishino-YamamotoECAL_2018}

\subsection{Derivatives of local parameters}\label{DerOfLoPa}
Derivatives of local parameters are obtained as follows.
\noindent
{\bf [1] Derivative of $\varrho$}\\
\begin{eqnarray*}
&& \frac{\partial \varrho_a}{\partial {\mib R}_a}   
    =  \sum_{b (\ne a)} \frac{\partial \varrho^a}{\partial R_{ab}} \cdot \frac{{\mib R}_{ab}}{R_{ab}} , \ \ \  
   \frac{\partial \varrho^a}{\partial {\mib R}_b}
    =  -\frac{\partial \varrho^a}{\partial R_{ab}} \cdot \frac{{\mib R}_{ab}}{R_{ab}}  \label{varrho-der2} \\
&& \frac{\partial \varrho^a}{\partial R_{ab} }  
 =  -\Lambda e^{-\Lambda R_{ab}}F_c(R_{ab}) \\
 & &  - \frac{1}{L}\cdot e^{(R_{ab}-R_0)/L} 
       \cdot e^{-\Lambda R_{ab}} F_c(R_{ab})^2 \ \ \ \ \label{varrho-der3}
\end{eqnarray*}

\noindent
{\bf [2] Derivative of $(1/r_s^a)^3$}\\
\begin{eqnarray*}
&&  \frac{\partial (1/r_s^a)^3}{\partial {\mib R}_{a}}
    =\frac{-3}{(r_s^a)^4}\cdot (r_s^a)^{m+1}\sum_{b(\ne a)} \Big(\frac{1}{R_{ab}}\Big)^{m+1} \cdot \frac{{\mib R}_{ab}}{R_{ab}}    \label{def-rsa-2}\\
&&  \frac{\partial (1/r_s^a)^3}{\partial {\mib R}_{b}}
    = -\frac{-3}{(r_s^a)^4}(r_s^a)^{m+1} \Big(\frac{1}{R_{ab}}\Big)^{m+1} \cdot \frac{{\mib R}_{ab}}{R_{ab}}    \label{def-rsa-3}
\end{eqnarray*}

\noindent
{\bf [3] Derivative of $\eta$}\\
\begin{eqnarray*}
&& \frac{\partial\eta^a}{\partial {\mib R}_{a}} 
 = \sum_{b(\ne a)} \frac{\partial \eta^a}{\partial R_{ab}} \cdot \frac{{\mib R}_{ab}}{R_{ab}}, \ \ \ 
\frac{\partial\eta^a}{\partial {\mib R}_{b}} 
= -\frac{\partial \eta^a}{\partial R_{ab}} \cdot \frac{{\mib R}_{ab}}{R_{ab}}, \nonumber \\
&& \frac{\partial\eta^a}{\partial R_{ab}}
 = \frac{m}{{R_{ab}}^{m+1}\Big(\sum_c \frac{1}{{R_{ac}}^m} \Big)^2}  \cdot
   \sum_{c (\neq b)} \frac{\log(R_{ac}/R_{ab})}{{R_{ac}}^m}.
\end{eqnarray*}

\subsection{Derivatives of overlap matrix} 
The overlap integrals are summarized in Slater-Koster table~\cite{S-K_Table} 
with direction cosines $\ell, m, n$. 
Consider an s-p interaction as an example; 
\begin{eqnarray*}
V_{s,x}({\mib R}) =\ell V_{sp\sigma}(R) .
\end{eqnarray*}
Its derivative is 
\begin{eqnarray*}
 \nabla_{\mib R}V_{s,x}=
  \frac{1}{R} \left(
    \begin{array}{lll}
        (-\ell^2+1)  \\
        -\ell m      \\
        -\ell n       
  \end{array} \right) V_{sp\sigma}
 +
  \ell  
   \left(
    \begin{array}{lll}
        \ell  \\
         m    \\
        n       
   \end{array} \right)\frac{{\rm d} V_{sp\sigma}}{{\rm d} R} .
\end{eqnarray*}
The first part is the angular-component derivative $\nabla(\ell)$ 
and the second part the radial-component derivative. 
One can prepare the table of the Slater-Koster coefficients in such way.
\subsection{Derivatives of TB Hamiltonian}
\begin{eqnarray*}
&& Z(\varrho, \eta)\equiv ((\varrho+\exp(g))(\tan(\eta)+\exp(h))+\exp(k)), \\
&& \frac{\partial \Delta(\varrho, \eta)}{\partial \varrho}  =  (\Delta(\varrho, \eta)-a) \Big[\frac{c}{(\varrho+\exp(e))} - \frac{d(\tan\eta+\exp(h))}{Z(\varrho, \eta)} \Big] , \\
&& \frac{\partial \Delta(\varrho, \eta)}{\partial \eta}  = -(\Delta(\varrho, \eta)-a) d \frac{(\varrho+\exp(g))(\tan(\eta)^2+1)}{Z(\varrho, \eta)} .
\end{eqnarray*}
The additional boundary condition $\delta\Delta(\eta)$ is also continuous and differentiable. 


\end{document}